\documentstyle[11pt,aaspp4]{article}

\newcommand{\mincir}{\ \raise -2.truept\hbox{\rlap{\hbox{$\sim$}}\raise5.truept
	\hbox{$<$}\ }}			
\newcommand{\magcir}{\ \raise -2.truept\hbox{\rlap{\hbox{$\sim$}}\raise5.truept
 	\hbox{$>$}\ }}	

\lefthead{Tr\`evese, Cirimele and De Simone}
\righthead{Matter Distribution in A~2319}

\begin{document}

\title{An X-ray and Optical Study of Matter Distribution\\
 in the  Galaxy  Cluster A~2319}

\author{D.Tr\`evese, G. Cirimele, and M. De Simone}
\affil{Istituto Astronomico, Universit\`a di Roma ``La Sapienza'',
via G.M. Lancisi 29, I-00161 Roma, Italy;\\
trevese@astro.uniroma1.it}

\bigskip
\begin{center}
Accepted for publication on Astrophys. J.
\end{center}

\begin{abstract}
A new analysis of velocity distribution, optical photometry
and X-ray surface brightness from ROSAT PSPC data of the galaxy
cluster A~2319 is presented. 
The temperature profile derived from ASCA data (\cite{Mar96})
is taken into account. A method to check the
hydrostatic model in the presence of a temperature gradient is
proposed. Consistency of the
hydrostatic isothermal model and the explanation of the
"$\beta$-discrepancy" are discussed. 
Galaxy and gas density profiles
of the main component A~2319~A are derived, allowing for
the effect of the secondary component A~2319~B.  
The inadequacy of a polytropic model, which would produce 
a binding mass decrease with respect to the isothermal $\beta$-model,
is discussed.
A simple interpolation of the temperature profile
provides instead an increase of the binding mass and  
a lower baryon fraction 
thus mitigating the ``baryon catastrophe''.
Assuming as typical the value $f_b \approx 0.2$,
 a comparison with the most recent
estimate of $\Omega_b^{nucl}$ implies for the cosmological parameter
$\Omega_o \mincir 0.4$.

\end{abstract}


\section{Introduction}

Detailed studies of the matter distribution in clusters of galaxies
provide important clues on the growth of condensations and 
the evolution of the Universe.
From X-ray observations it is possible to derive both the gas and the 
total binding
mass distributions, under the assumption of hydrostatic equilibrium.
Optical data, i.e. galaxy photometry and redshifts, combined with X-ray
observations allow to check the validity of the hydrostatic equilibrium
assumption and  derive the spatial distribution of the dark matter.
Most analyses in the past 
(\cite{JF84,CHM87,HGF88,HUG89,GER92,BHB92,DUR94,DJF95,CNT}) 
were based on the further 
simplifying assumption that the gas is isothermal, at least within about 1 
h$_{50}^{-1}$ Mpc, possibly with the exclusion of
a central cooling flow region (see e.g. \cite{FNC84,WJF97}).
This leads to the $\beta$-model (\cite{CFF76})
which predicts that the dynamical parameter 
$\beta_{spec}\equiv \frac{\mu m_p\sigma_r^2}{k T}$, representing
the ratio between the energy per unit mass of galaxies and gas,
equals the morphological parameter $\beta_{fit}$ defined by the fit of the
gas density distribution with a King profile. 
The observations show that on average $\beta_{fit} < \beta_{spec}$ 
(\cite{Sar86,EVR90}). 
However Bahcall and Lubin (1994) ascribed this "$\beta$-discrepancy"
to the underestimate of the slope of the galaxy density profile, 
appearing in the hydrostatic equilibrium equation,
rather than to a failure of the model. In their X-ray--Optical analysis of 
a sample
of Abell clusters, Cirimele, Nesci, \& Tr\`evese (1997) (CNT) found that
$\log \rho_{gas}= \beta_{XO} \log \rho_{gal} + C$ in a wide range of
densities, as predicted by the hydrostatic isothermal equilibrium.
This allows to define, for each cluster, a morphological 
parameter $\beta_{XO}$,
independent of any analytical representation of $\rho_{gas}$ and $\rho_{gal}$.
A comparison of $\beta_{XO}$ with $\beta_{spec}$ supports the explanation 
of the "$\beta$-discrepancy" suggested by \cite{BL94} and the 
consistency of the "$\beta$-model", at least for several galaxy clusters
of regular and relaxed appearance.
The gas and binding mass distributions thus derived provide a typical
value of the baryon fraction
$f_B$, of the order of 0.2 within 1-2 h$_{50}^{-1}$ Mpc
(\cite{CNT,evr97,EttFab99},Mohr et al., 1999). This relatively high
value, compared 
with the results of primordial nucleosynthesis calculation (\cite{Wal91,OS95}
,but see \cite{BuTy98})
implies that either the cosmological parameter is smaller than unity
(\cite{Whal93}), or  $f_B$ is not representative of the cosmic value, and
galaxy clusters are surrounded by extended halos of non baryonic dark matter
(\cite{WF95}). The latter hypothesis raises the problem of
understanding the mechanisms of a large scale baryonic segregation.

However, recently ASCA data have provided direct evidences of gas temperature 
gradients in the outer regions of several galaxy clusters 
(\cite{Arn94,Mal94,Mal96,IKE96,Mar96,Mal98}).
According to \cite{Mar96}, in the outer regions some clusters show 
a polytropic index even greater than 5/3, which is inconsistent 
with the hydrostatic equilibrium conditions. 
According to \cite{EttFab98}, a systematic difference between 
the electron and the proton temperatures cannot explain the
inconsistency, and a real departure from hydrostatic equilibrium must 
happen in some cases.
Even disregarding these extreme cases, once the temperature profiles are
available it is worth: i) to check how far from the cluster center
the hydrostatic condition can be assumed, specially if strong temperature
gradients are present; ii) to estimate the correction to the total mass and 
baryon fraction implied by non-isothermality.
Moreover some clusters show an anomalously high value of the dynamical
parameter $\beta_{spec}$, possibly suggesting deviations from the equilibrium
conditions and requiring a detailed analysis of the velocity distribution.
The galaxy cluster A~2319 has been extensively studied in the past,
so that many galaxy redshifts are available, ROSAT PSPC images can
be retrieved from the public archive and a temperature profile based on ASCA
data has been published by \cite{Mar96} (see however \cite{MOL98},
\cite{MOL99}).

In the present work we combine these data  with the F band  photometry of
galaxies (\cite{Tal92}), and compare the gas and galaxy 
density distributions.
We generalize the definition of the morphological parameter $\beta_{XO}$
to verify the hydrostatic equilibrium conditions.
The analysis suggests the consistency of the hydrostatic model in the presence
of a temperature gradient. 
Thus we discuss the mass distribution as obtained by 
adopting a polytropic model
or a simple parabolic representation of the temperature profile, and
we compare the resulting baryon fraction
with the limits provided by the standard nucleosynthesis calculations, 
then deriving constraints on  
the large scale baryon segregation and the cosmological parameter $\Omega_o$.

We use $H_o= 50 ~h_{50} ~km s^{-1} Mpc^{-1}$.

\section{The galaxy distribution}

The galaxy cluster A~2319 has been studied by several authors
in the radio , optical and X-ray bands.
It is classified as a BM type II-III and as a richness 1, RS-type cD
cluster (see \cite{ACO}). 
The galaxy velocity dispersion $\sigma\approx$1800 km s$^{-1}$ 
is particularly high. However,
Faber and Dressler (1977),
on the basis of 31 galaxy spectra already suggested that A~2319 is actually 
two clusters nearly superimposed along the line of sight: the main component
A~2319~A  with an average redshift $\bar{v}_A$=15882 and a velocity 
dispersion $\sigma_A=873^{+131}_{-148}$ km s$^{-1}$ and the second component 
A~2319~B, located about $8^\prime$ NW with $\bar{v}_B$=19074 and  
$\sigma_B=573^{+120}_{-149}$ km s$^{-1}$.

More recently Oegerle, Hill, and Fitchett (1995)(OHF)
measured several new redshifts, 
applied the "$\delta$-test"of Dressler and Shectman (1988)
to locate the A and  B components, and assigned
the 139 galaxies of known redshift to the component A and B 
(or to the background/foreground)
on the basis of their position and redshift, empirically trying to keep
gaussian the velocity distribution of A~2319~B.
They found N$_A$=100 and  N$_B$=28 galaxies in the two components with
$\bar{v}_A$=15727 km s$^{-1}$, $\sigma_A$=1324 km s$^{-1}$ 
$\bar{v}_B$=18636  km s$^{-1}$ $\sigma_B$=742 km s$^{-1}$ respectively.

To assign individual galaxies to the A and B components we  adopted the 
results  of OHF to obtained a first order
estimate of the cluster positions, average radial velocities $\bar{v}^{(i)}$, 
and velocity 
dispersions $\sigma_{(i)}$, and we computed the relevant core radii 
$R_c^{(i)}$ of the two components, where i=A,B identifies the component.
Then we assumed the following probability distributions of galaxies
respect to radial velocities v and projected distance b from the
relevant cluster center:

\begin{eqnarray}
P_i(b,v) & \equiv & \frac{N_i}{N_A+N_B} \cdot f_i(b) \cdot g_i(v)  \\
f_i(b) & = & \left\{ \frac{2\pi R_c^{{(i)}^2}}{1-\beta_{(i)}}
\left( 
\left[ 1+\left(\frac{b_{max}^{(i)}}{R_c^{(i)}}\right)^2\right]^{(1-
\beta_{(i)})}-1\right)\right\}^{-1}
\left[ 1+\left(\frac{b}{R_c^{(i)}}\right)^2\right]^{-\beta_{(i)}}  
\nonumber \\
g_i(v) & = & \frac{1}{\sqrt{2\pi \sigma_{(i)}^2}}
\exp\left[ -\frac{(v-\bar{v}_{(i)})^2}{2\sigma_{(i)}^2}\right]  \nonumber
\end{eqnarray}
\noindent
where $N_{(i)}$ the first order estimate of the number of galaxies of the 
relevant component, and the $b_{max}^{(i)}$ is the radius of the circle 
containing the $N_i$ observed galaxies. 
This  simple parameterization is independent of any assumption about 
the distance along the
line of sight of the two clusters and their relative motion.
Each galaxy is then  assigned to the component 
of higher probability.
The 99\%  confidence volumes  are also 
considered for each cluster, and galaxies outside these volumes are 
assigned to the background or foreground.
We obtain the new values  N$_A$=96 and  N$_B$=24 
$\bar{v}_A=15891 km s^{-1}$, $\sigma_A=1235 \pm 90 km s^{-1}$,
$\bar{v}_B=18859 km s^{-1}$, $\sigma_B=655 \pm 97 km s^{-1}$.
Since the resulting velocity distributions of the two components
do not show strong deviation from gaussian distributions, the 
reported uncertainties $\sigma_{\sigma_A}$ and $\sigma_{\sigma_B}$,
on $\sigma_A$ and $\sigma_B$ respectively, have been computed as
$\sigma_{\sigma_i}^2=\sigma_{(i)}^2/[2(N_{(i)}-1)]$ ~i=A,B. 
The effect of membership uncertainty can be evaluated as follows.
Given a sample of N galaxies with average velocity $\bar{v}$ and velocity 
dispersion $\sigma$, the addition of k galaxies with velocity
$v=\bar{v}+\delta$ produces a new velocity dispersion 
$\sigma'^2=(N \sigma^2 + k \delta^2)/(N+k-1)$.
Therefore, to increase $\sigma_A$ by more than $2\sigma_{\sigma_A}$ it is
necessary to include in the sample A more than k=2 galaxies with a
recession velocity exceeding 
$\bar{v}_B+2 \sigma_B\simeq \bar{v}_A + 3.46 \sigma_A $ .

Although on the sole basis of the "$\delta$ test" there is a 10\%
probability that  A2319B is not a physical association, the strong
clustering of large $\delta$ values in a region (see OHF fig.5)
corresponding to enhanced X-ray emission suggest that it
is a physical entity.  Moreover, the analysis of bound orbits of the
two components A~2319~A and A~2319~B led \cite{OHF} to the conclusion
that "there is a reasonably high probability that these clusters are
not bound and will never merge".  The latter conclusion is supported
by the discussion of FGB who compare the X-ray images with
simulations of cluster collisions (see section 3).
The above considerations suggest and legitimate the assumption, 
adopted in the following, that the two clusters are separate entities.

We have added to the spectroscopic information the F band photometry of
A~2319, obtained 
by \cite{Tal92} from microdensitometric scans of a Palomar 48 inch 
Schmidt plates, as part of a systematic study of the morphology and 
luminosity functions of galaxy clusters (\cite{TCF92,FTCH95,tca96,Tal97}).
Due to the low galactic latitude (b$\simeq 13^\circ$) the field of
A~2319 is very crowded and the automatic star/galaxy classification 
is difficult. Thus we have revised the classification and recovered some 
misclassified object. 

To focus our attention on the main component A,
we reduced the effect of the B component excluding from
our sample  all the galaxies classified  as B (or background/foreground). 
We adopted a fixed center
$(\alpha=19^h 21^m 11.8^s  , \delta=+43^o 56' 39'' (J2000))$,
derived from the centroid of X-ray emission 
as computed in a small (2 arcmin) circle around the intensity peak.
This point is identified with the center of a spherical structure which we 
assume to represent the A component.
We chose a magnitude limit $m_F=m_3+2$=16.33 mag and the resulting fraction 
of galaxies without measured redshift is 0.23.
Thus, the fraction of galaxies without measured redshift and belonging to the B
component is of the order $0.23 N_B/(N_A+N_B)$, i.e. 4\%, and is not expected 
to affect significantly the galaxy density profile.
We fitted with a maximum likelihood algorithm the unbinned
galaxy distribution  using both a King profile 
$\sigma_{gal}(b)=\sigma_0 \cdot (1+(b/r_c)^2)^{-\kappa} + \sigma_b$ 
and with a de Vaucouleurs profile
$\sigma_{gal}(b)=\sigma_0 \cdot exp(- 7.67(b/r_v)^{\gamma_g}) + \sigma_b$,
where the background counts $\sigma_b$, $r_v$, $\gamma_g$, $r_c$ and
$\beta$ are free parameters, while $\sigma_0$ is determined by the
normalization to the total number of observed galaxies and
$b$ is the projected distance from the cluster center.  
The Kolmogorov-Smirnov (KS) test has been applied in both cases
and the results are reported in Table 1, where the errors reported 
represent one-sigma uncertainties derived from Monte Carlo simulations
described in section 4,
and $P_{KS}(>D)$ is the probability 
of the null hypothesis that deviations larger than D are produced by random 
noise. 

\placetable{tbl-1}

The surface density and the fitting profiles are shown in Figure 1.
In this case
the King profile has a slightly higher probability and will be adopted in the
following to derive the volume distribution by numerical inversion. 
However, the differences between the two fitting curves, specially for 
$b>0.1 h_{50}^{-1} Mpc$, cannot affect significantly the subsequent 
discussion of the hydrostatic equilibrium conditions.

\placefigure{fig1}

To obtain the total luminosity of the cluster, $L_{tot}(r)$,
we fitted with a \cite{Sche76} function the unbinned luminosity distribution
excluding the brightest member,
using a maximum likelihood algorithm, adopting a constant $\alpha=-1.25$ 
and $M^{\ast}$ as a free parameter, as in \cite{tca96}. 
This gives:
\begin{equation}
 L_{tot}(r) = 10^{-0.4(M_F^\ast + 28.43)} { { \Gamma(2+\alpha)\cdot
4\pi}\over{ \Gamma (1+\alpha,{ {L_{lim}\over{L^\ast}} })}}\cdot
\int_0^r \rho_{gal}(r')r'^2 dr'
\end{equation}
where  $L_{tot}$ is expressed in units of $10^{13} L_{\odot}$, $r$ in $kpc$ and
$L_{lim}$ is the limiting luminosity corresponding to the
magnitude limit ($M_F=-21.56$ mag) of the galaxy sample adopted to derive 
$\rho_{gal}$. The value of $M^{\ast}$ changes by less than 1\%
considering a sample which includes the galaxies of A~2319~B.
The galaxy mass $M_{gal}(r)$ is then
obtained from the  total luminosity
assuming an average mass-to-light ratio
$M/L_R=3.32 \pm 0.14 M_{\odot}/L_{R \odot}$ from \cite{vdma91}, 
adopting $F \simeq R$ for bright ellipticals (see \cite{Lug89})
and computing $F_{\odot}$ from the relation V-F=0.40 (B-V) (\cite{Kron80}).

\noindent

To estimate the virial mass we have evaluated the r.m.s. velocity
dispersion $\sigma_r$ in four concentric rings each containing 1/4 of
the galaxies of known redshift of A~2319~A. The four values are
1148 km s$^{-1}$, 1415 km s$^{-1}$, 1327 km s$^{-1}$, 1193 km s$^{-1}$
with a r.m.s uncertainty of about 200 km s$^{-1}$. 
Thus in the following we assume a
constant dispersion, derived from the entire A~2319~A sample,
$\sigma_r=1235 \pm 90 $km s$^{-1}$

The resulting virial mass  
$M_V=3\pi b_G \sigma_r^2/2G$ 
is $M_V=2.89 \times 10^{15} h_{50}^{-1}{\cal M}_\odot$, namely only 2\% less 
than the value given by OHF, since the decrease of $\sigma_r^2$
is almost entirely compensated by a slight increase of the projected
virial radius $b_G=2 \langle 1/b \rangle^{-1}$ (\cite{Sar88}), 
which in our case is $b_G=1.736 h_{50}^{-1}$ Mpc.

\section{The gas distribution}

From the ROSAT public archive we extracted the available Position
Sensitive Proportional Counter (PSPC) images corresponding to two
observations of A~2319, on March 1991 and
November 1992, which cover a 128x128 arcmin$^2$ field with pixel size
of 15x15 arcsec$^2$ and an effective resolution of about
$25^{\prime\prime}$ FWHM, in the energy band
0.5-2.0 keV.
The exposure maps (\cite{Snal92,Plal93}), providing the effective exposure time
of each pixel, are also available at the ROSAT public archive for each 
observation.
We divided each image for the relevant exposure map and combined
the two images with weights proportional to the maxima of the exposure
map (1514.6 s and 3200.8 s respectively).
The resulting image is shown in Figure 2.  
 
\placefigure{fig2}
 
Feretti, Giovannini \& B\"ohringer (1997) (FGB) discuss 
the radio structure of A2319, which shows a powerful radio halo.
They also analyze two
substructural features in the X-rays,
one corresponds to the E-W elongation in the
very center of the A component, detected in the image obtained with 
the ROSAT High Resolution Imager, and is interpreted by FGB as an evidence
of a recent merging process, likely providing the energy for
the radio halo . This feature is confined within the inner
5 arcmin and does not affect the analysis of the hydrostatic equilibrium,
particularly in the region where the temperature is not constant, i.e. 
for $r > 5$ arcmin.
In the outer region the cluster structure is rather regular, as for
most cD clusters, except for an elongation in the direction of the B
component.  According to FGB, the B component is in a pre-merger
state and has not yet proceed far enough to disturb the bulk of the gas,
as can be argued from a comparison with the cluster collision simulation by
\cite{Schimu93}.

Thus we analyzed A~2319~A as a separate entity, as discussed in section 2.
As a first approximation we ignored the presence of the B
component, we assumed spherical symmetry and, as in CNT, we derived
the volume density $\rho_{gas}(r)$ of the gas by both numerical
inversion of the projection equation, and by fitting the observed
surface brightness with a "$\beta$-model" (\cite{CFF76})
$I(b)=I_0 [1+(b/r_c)^2]^{-3\beta +1/2}+I_b$, obtaining consistent results.
The results are also consistent with FGB and with a more recent analysis
of \cite{Moh_al99}.
Then we applied the same procedure after the exclusion of the northern
half ($\delta > 43^o 56' 39'' (J2000)$) of the image, to eliminate the effect 
of the B component.  
We have evaluated the constant background value 
$I_b=8.86\times 10^{-4}$ cts s$^{-1}$ arcmin$^{-2}$
in the  region $b > 3 h_{50}^{-1}$ Mpc.
The fitting parameters  
are reported in Table 2, together with the
central proton density $n_o$. The one-sigma uncertainties have been evaluated
by Monte Carlo simulations, described in section 4.
The central proton density is obtained from the relation:
\begin{equation}
I_0 [1+(b/r_c)^2]^{-3\beta +1/2}= 
EM \cdot \int_{\nu_{min}}^{\nu_{max}} \Lambda(\nu,T) d \nu
\end{equation}
where $\Lambda(\nu_{rest},T)$ is the rest-frame cooling function 
corrected for the cosmological
dimming factor $(1+z)^4$ and computed with the code of \cite{meweal}
with 0.3 solar abundance corresponding to $\frac{n_e}{n_p}=1.2$
, $\nu_{min}$ and $\nu_{max}$
define the observing band in the rest-frame and 
$EM \equiv 
\int_0^{\infty}(\frac{n_e}{n_p}) n_o^2 (1+\frac{r^2}{r_c^2})^
{-{\frac{3}{2}\beta}} dl$~
(see \cite{Sar88}), 
and  correction for  absorption  with 
$N_H = 8.89\cdot 10^{-20} ~cm^{-2}$ (\cite{stark}) has been applied to 
$I_o$.

\placetable{tbl-2}
 
Again our results are consistent with the analysis of FGB. 
In particular we
also find a smaller value of the core radius $r_c$ of the A
component when the northern half is excluded to reduce the effect
the B component.
The observed surface brightness profile and the fitting $\beta$-model
are shown in Figure 3.  
 
\placefigure{fig3}

For consistency with the optical analysis, where we eliminated  the 
galaxies assigned to the B component, in the following
we adopt the fit obtained after the exclusion of the northern half
of the cluster image.
The corresponding gas density profile $\rho_{gas}(r)=\rho^o_{gas}
[1+(b/r_c)^2]^{-3\beta/2}$ is obtained assuming a constant temperature
equal to the emission-weighted temperature $T_X=10.0 \pm 0.7 keV$ 
(\cite{Mar96}).
The non-isothermality does not significantly affect the results
(\cite{Mal96}), due to the weak dependence of the emissivity on
temperature, in the adopted band (see next section).
The gas density profile has been computed also by a non parametric
numerical deprojection as in CNT, obtaining consistent results.

\section{ Hydrostatic model, mass distributions and baryon fraction}

As pointed out in CNT, it is possible to check the hydrostatic equilibrium
condition in a non-parametric way, by a direct comparison of the density
profiles of gas and galaxies. 
Assuming spherical symmetry, the equilibrium condition implies 
(\cite{BL94}):
\begin{equation}
{ {\mu m_p\sigma_r^2}\over{k T}} =
{  
{ d \ln \rho_{gas}(r)/d \ln r + d \ln T/d \ln r} \over 
{ d \ln \rho_{gal}(r)/d \ln r + d \ln \sigma_r^2/d \ln r + 2 A}  
}
\end{equation}
where $\sigma_r$ is the radial galaxy velocity dispersion, 
$A=1-\left({\sigma_t \over \sigma_r}\right)^2$ measures the 
anisotropy of the velocity distribution and $\mu$ is the average molecular
weight which we assume equal to 0.58 (\cite{ES91}).
For constant $\sigma_r$ and $T$, and
$A=0$, this implies $\rho_{gas} \propto \rho_{gal} ^{\beta}$
where $\beta_{spec}={ {\mu m_p\sigma_r^2}\over{k T}}$ is a constant 
representing the ratio between the energy per unit mass in the galaxies
and the gas respectively (\cite{CFF76}).
In this case it is possible to define the morphological parameter
$\beta_{XO}\equiv d \ln \rho_{gas}(r)/d \ln \rho_{gal}$
to be compared with $\beta_{spec}$ which is obtained from the spectroscopic
observation of galaxies and the gas temperature derived from X-ray spectra.
As already discussed in CNT, the very existence of a wide range of
densities where $\beta_{XO}$ is constant supports, in many cases,
the validity of the 
isothermal model. However in the presence of a temperature gradient, as in 
the case of A~2319~A, both $\beta_{spec}$ and $\beta_{XO}$ depend on radius
and the equilibrium equation reads:
\begin{equation}
\beta_{spec}=\beta_{XO}+\beta_{TO},
\label{bb}
\end{equation}
where $\beta_{TO}\equiv d \ln T/ d \ln \rho_{gal}$ and we still assume that
the galaxy velocity distribution is isotropic and $\sigma_r$ is constant.
To check the validity of equation~\ref{bb}  we  used the temperatures
obtained by \cite{Mar96}.
We computed $\beta_{spec}$, $\beta_{XO}$ and $\beta_{TO}$ at three 
projected radii of 4, 10 and 20 arcmin,
corresponding to the boundaries of the four rings
whose temperatures are given by \cite{Mar96}.  With
the constant velocity dispersion $\sigma_r=1235\pm 90$km s$^{-1}$
derived in section 2, $\beta_{spec}$ ranges from 0.86 $\pm$ 0.15 in the
first point, to 1.52$\pm$ 0.47 in the outermost point, i.e. 
the ratio between the energy per unit
mass in gas and galaxies  is close to unity within a central
isothermal region of about 0.5 h$_{50}^{-1}$ Mpc, while it decreases 
in the outer regions. 

Figure 4 shows $\ln \rho_{gas}$, obtained by numerical deprojection, 
versus $\ln\rho_{gal}$. 
From the slope of the curve is possible to derive a value
of $\beta_{XO}(r)$ at each radius.  

\placefigure{fig4}

The values of $\beta_{XO}(r)$ range from $0.59\pm 0.06$
in the first ring, to $0.64\pm0.06$ in the outermost ring.
The one-sigma uncertainties are evaluated from  Monte Carlo simulations
described below. 
We also
computed $\beta_{TO}(r)$, which is close to zero in the cluster center
and increases up to 0.58$\pm0.44$ at r$\approx$1 h$_{50}^{-1}$ Mpc.
The one-sigma uncertainties are derived from errors 
on the temperature values reported in Figure 2 of \cite{Mar96}.  Figure 5
shows $\beta_{spec}(r)$ as a function of the RHS of equation~\ref{bb}, 
($\beta_{XO}+\beta_{TO}$).

\placefigure{fig5}

We stress that the relation between the two quantities on the x and y
axes is not automatically implied by their definition, since the
galaxy density $\rho_{gal}(r)$ only appears in the RHS and is
observationally independent of the quantities in the LHS.

The value of $\beta_{spec}(r)$ is slightly larger than $\beta_{XO}+\beta_{TO}$.
This could indicate that $\sigma$ is still slightly overestimated
due to the presence of the background component A~2319~B.
However the intrinsic
statistical uncertainty does not allow this level of accuracy.
Furthermore, even small deviation from the spherical symmetry, 
or from isotropy of the velocity distribution, could have 
comparable effects.

All we can safely say is that there is an increase of
$\beta_{spec}$ versus  ($\beta_{XO}+\beta_{TO}$),
which is consistent with a straight line
of unit slope. 
Thus, within the present uncertainties the data are consistent
with equation \ref{bb},
namely with validity of  the hydrostatic equilibrium, 
also in the outer region of the cluster where the temperature declines.

The accuracy  of this type of check, as applied to a single cluster, is
limited by various factors. Although future X-ray data will provide
much higher signal-to-noise, the uncertainty on galaxy density
profile is intrinsically limited by Poisson noise on galaxy counts.
Moreover subclustering and unknowable deviations from spherical symmetry  
will always produce an uncertainty on the galaxy density deprojection.
 
Nevertheless, the systematic application of the method described, to all
the cluster with measured temperature distribution (\cite{Mal98})
will likely provide statistical indication on the validity of, or the
deviation from, the equilibrium conditions in the outer parts of
galaxy clusters.

Assuming that our results indicate the validity of hydrostatic equilibrium,
we can derive the distribution 
of the total binding mass $M_T(r)$ of A2319A:
\begin{equation}
M_{tot}(r)=-{{kT}\over{\mu m_p G}}\left({ {d \ln \rho_{gas}(r)}\over{d \ln r}}
+ { {d \ln T}\over{d  \ln r} } \right) r
\label{mtot}
\end{equation}

The  slopes of the temperature profiles are crucial
in establishing whether the non-isothermality causes an increase or decrease
of the mass estimate, with respect to the isothermal $\beta$-model.

In fact, from equation~\ref{mtot},
indicating with  $M_{tot}^{isot}(r)$ the mass 
derived by an isothermal $\beta$-model with  temperature $T_{isot}$
the fractional change in the mass estimate is:
\begin{equation}
\Delta(r) \equiv \frac{M_{tot}(r)-M_{tot}^{isot}(r)}{M_{tot}^{isot}(r)}=
\frac{T(r)-T_{isot}}{T_{isot}}+
 \frac{T(r)}{T_{isot}} \frac{ d \ln T}{d \ln \rho_{gas}}
\label{delta}
\end{equation}
where the two terms on the r.h.s. of the equation can be of the same order.

It is customary to adopt a polytropic gas distributions as the simplest
analytic representation of a non-isothermal gas in hydrostatic equilibrium.
In this case $T \propto \rho_{gas}^{\gamma-1}$, with the polytropic index
$\gamma$ ranging from unity to 5/3 for an isothermal or adiabatic
equilibrium respectively. The polytropic model
implies (\cite{CHM87}):
\begin{eqnarray}
&&\rho_{gas}(r)=\rho^o_{gas} [1+(r/r_c)^2]^{-\delta},~~~~~ 
T(r)=T_o [1+(r/r_c)^2]^{-\alpha}, \\
&&\delta=\frac{3\beta/2}{1+\eta (\gamma-1)/2},~~~~~
\alpha=\delta (\gamma-1) \nonumber
\label{politro}
\end{eqnarray}

\noindent
where $\eta \equiv d \ln \epsilon / d \ln T$ and $\epsilon$ is the emissivity 
of the gas, integrated in the adopted band.
We can fit the temperature profile of A~2319 provided by \cite{Mar96},
neglecting the effect of projection and adopting $\eta \simeq -0.2$,
to determine the polytropic index $\gamma$. 
Notice that,
due to small value of $\eta$, the result is not significantly different
if we simply assume $\delta=3\beta /2$ and $\eta=0$ in the above expressions.
The slope  $\left| dT/dr \right|$ of the temperature profile reaches a maximum
at $r/r_c=-\alpha+\sqrt{\alpha^2+1}$ and progressively decreases in the 
outer regions.

As a result, for $T_{isot}=T_o$, the quantity $\Delta(r)$ is 
positive 
only for $r/r_c<x_{\gamma}\equiv\sqrt{\gamma^{1/\alpha}-1}$. 
This limit decreases for increasing
$\beta$ and $\gamma$, e.g. $x_{\gamma}\simeq 0.78$  for $\beta=1$,
$\gamma=5/3$ and $\eta=-0.2$,
while $x_{\gamma}=\sqrt{e^2-1}\simeq 2.53$ for $\beta=1/3$ and $\gamma=1$.
We stress that $\Delta(r)$ is always negative at large radii, and this
implies an enhancement of the ``baryon catastrophe''.   
The best fit  value is $\gamma=1.091$ and the resulting temperature profile
is shown in Figure 6. 

\placefigure{fig6}

The quality of the fit is quite poor, and there is a probability of
91\% that the deviations are non random.  The increment of the mass in
the central region, with respect to isothermal model, and the
enhancement of the baryonic catastrophe obtained with a polytropic
model are a mere artifact of the particular shape of the polytropic
temperature profile, which is not a good representation of the data,
at least in the case of A~2319~A.

More specifically,  the data seem to 
indicate an almost isothermal central region and an increasing slope
of the temperature profile for increasing radius.

This behavior is consistent with a change of the polytropic index with radius
from an isothermal 
($\gamma=1$) towards an adiabatic ($\gamma=5/3$) hydrostatic equilibrium
(see \cite{Sar88}).
A fit with the law $T(r)=T_o-a r^2$ is  shown in Figure 6. In this case
the  probability is P($>\chi^2$)=0.99. 

Figure 7 shows $M_{tot}^{isot}(r)$, corresponding to the isothermal model, 
and $M_{tot}(r)$ as obtained with both
the polytropic model and the quadratic interpolation
of the temperature profile. On the basis of the above discussion we assume 
the latter as the best representation of the mass distribution. 

\placefigure{fig7}

In the same figure, galaxy and gas masses,
$M_{gal}(r)$ and $M_{gas}(r)$ are also shown. The latter is computed both in 
the isothermal approximation and using the temperature profile:
the result is only weakly dependent on the temperature changes.

$M_{gas}(r)$ is steeper than 
$M_{gal}(r)$ and $M_{tot}(r)$, which  have more similar slopes.
As a consequence the gas mass dominates over
the galaxy mass at large radii.
These results are consistent with previous findings of CNT.

The statistical uncertainties have been evaluated 
by Monte Carlo simulations of the entire reduction process.
For the X-ray data, starting from a "$\beta$-model" corresponding to the 
fitting parameters,
we generated 500 random sets representing the photon counts in each radial
ring with Poisson noise, and we  extracted 500  random  background values,
with a standard deviation  estimated from intensity fluctuation
of the surface brightness  outside $3 h_{50}^{-1}$ Mpc where the 
background value 
has been measured. Then we  fitted with a  "$\beta$-model" each intensity
profile, obtaining the statistical distribution of the fitting parameters.
Finally we extracted 500 values of the temperature in 
each of the four rings corresponding to the data of \cite{Mar96},
with the relevant standard deviations,  and we fitted
the temperature profile with a parabolic law.
Then we applied to the simulated data the same algorithms applied to real data
for the evaluation of the mass profiles.
This procedure allows to define a one-sigma confidence interval for 
the gas mass $M_{gas}$ and for the total mass $M_{tot}$ as a function of 
radius.
A similar procedure was adopted for the galaxy distribution.
Then we extracted, for each cluster simulation, a random value of $M/L_{F_\odot}$
from a gaussian distribution with a mean value and a standard deviation 
corresponding obtained from \cite{vdma91}. 
The one-sigma confidence intervals are reported as shaded areas in Figure 7.

We define the luminous mass as the sum of the gas mass and
the galaxy mass as deduced using an average  stellar
mass-to-light ratio (see section 2): 
$M_{lum}=M_{gas}+M_{gal}$. The above results imply that the dark matter
$M_{dark}=M_{tot}-M_{lum}$ has a distribution similar to $M_{gal}(r)$.
This provides a constraint on the mechanism of galaxy and cluster formation.

Since an unknown fraction of the dark matter is baryonic, $M_{lum}/M_{tot}$
represents a lower limit on the baryon fraction $f_b$.
Figure 8 shows this lower limit as a function of radius,  as computed 
in the isothermal approximation and taking into account the
temperature gradient by the polytropic model and by the quadratic 
interpolation. 
We can compare our  estimate of $f_b$ with the results of
\cite{Moh_al99} who gives the $f_b$ values 
$f_b=0.213\pm0.004$ and $f_b=0.297\pm0.027$, at $1 h_{50}^{-1}$ Mpc and
at $1.91 h_{50}^{-1}$ Mpc respectively, the latter distance representing 
$r_{500}$, namely the radius within which the mean density is 500 times
the critical density $\rho_{crit}=3 H_o^2/8\pi G$.
Our isothermal values  $f_b=0.180\pm0.017$ and $f_b=0.252\pm0.045$   
are slightly lower, because were derived by a fit of 
the southern part of the X-ray  image, to exclude the effect
of the B component.   

\placefigure{fig8}

At 2 $h_{50}^{-1}$ Mpc $f_{b}$, as computed with the quadratic 
interpolation, becomes respectively 66 \% and 56 \%
of the  isothermal and polytropic values,
thus mitigating the "baryon catastrophe".

Assuming $f_b \simeq 0.2$ as typical of galaxy clusters,
the residual  discrepancy between $\Omega_b=f_b \Omega_o$
and the corresponding value derived from nucleosynthesis calculations
$\Omega_b^{nucl}\simeq 0.076 \pm 0.004 h_{50}^{-2}$ (\cite{BuTy98}),
can be reconciled assuming 
$\Omega_o < \Omega_b^{nucl}/f_b\mincir 0.4$.

\section{Summary and Conclusions}

We have performed a new analysis of the Abell cluster A~2319, and assigned
the individual galaxies to the A~2319~A and A~2319~B components by an
objective criterion taking into account both position and redshift.
The resulting velocity dispersion of the A component is slightly
smaller, but consistent with the previous determination of OHF.

We have obtained photographic F band photometry of the cluster
galaxies, which allows us to construct the cluster luminosity function and
galaxy density profile.

We have analyzed archival ROSAT PSPC images separating the A component
on the basis of the optical information, and we have obtained a gas
density profile of the A component. The result is consistent with 
recent studies of Feretti, Giovannini and B\"ohringer (1997), Mohr et al. 
(1999).

Since, according to \cite{Mar96}, A~2319~A shows a radial gas
temperature decrease, we have generalized the method introduced by
Cirimele, Nesci \& Tr\`evese (1997) CNT, in order to check the validity of
the hydrostatic equilibrium in the case of a non isothermal gas.

We have derived the total mass profile $M_{tot}(r)$ through the non
isothermal hydrostatic equation adding new evidence in favor of the
results of CNT that the total mass and the galaxy mass have
similar radial distributions, more concentrated with respect to the gas
component.

Polytropic models imply smaller masses at large radii, respect
to the isothermal model,  i.e. a higher value of the baryon fraction.
Thus, the use of a polytropic model would enhance the baryon catastrophe.

However we have shown that the polytropic model is
inconsistent with the observed temperature profile, at least in the specific
case of A 2319 A.  A parabolic representation of the temperature
profile gives, instead, a total mass larger than computed in the
isothermal approximation, mitigating the baryon catastrophe.

In any case,   $f_b$ is larger than
$\Omega_b^{nucl}\simeq 0.076 \pm 0.004 h_{50}^{-2}$, resulting from
nucleosynthesis calculations and recent measures of the deuterium to
hydrogen ratio (D/H), in high resolution studies of the $Ly{\alpha}$
forest (\cite{BuTy98}).
Under the assumption that the value of $f_b\approx 0.2$, 
found for A 2319 A, which is consistent with other estimates
(\cite{evr97}, CNT, \cite{EttFab99}, Mohr et al., 1999)
is typical of galaxy clusters, it is possible to  derive the following 
conclusions. If the $\Omega_o=1$
assumption is kept, then  the baryonic fraction within galaxy  clusters 
is not representative of the cosmic value and clusters must be surrounded 
by dark matter halos (\cite{WF95}). 
If, on the other hand, beyond $r\approx 2 h_{50}^{-1}$ Mpc the 
material has not yet fallen into the cluster, as infall models
indicate, then $f_b$ is 
representative of the cosmic value and the
cosmological parameter must be $\Omega_o=\Omega_b^{nucl}/f_b$
(\cite{Whal93} and refs. therein), i.e. in our case 
$\Omega_o\mincir 0.4$, where the uncertainty on the latter value
depends on the intrinsic spread of the baryon fraction, in
the presently observable volumes around galaxy clusters.

Recently Markevitch et al. (1998) have collected temperature profiles for 
30 galaxy clusters based on ASCA observations.
However, only a minority of clusters is regular enough, specially in
the outer regions, to allow an X-ray and optical check 
of the equilibrium conditions as suggested in the present paper.
This limits the accuracy and the reliability of a 
``measure'' of the cosmological parameter
$\Omega_o$, based on a comparison between the baryon fraction $f_b$ and
$\Omega_b^{nucl}$. Moreover future cosmic microwave background
experiments are expected to provide tighter constraints on $\Omega_o$
(\cite{MAND95}) as compared with the results derived from the
analysis of the matter distribution in galaxy clusters.

On the basis of this new estimate of $\Omega_o$, 
once the equilibrium conditions of the non-isothermal
regions are verified by the systematic application of the analysis
outlined in the present paper , it will be possible to extend the
(otherwise questionable) estimates of the general distribution of
luminous and dark matter based on the hydrostatic model, to the outer 
regions of  galaxy clusters
of a statistical sample, thus providing new constraints for
the models of cluster formation an the physics of large scale baryon
segregation.

We are grateful to the anonymous referee and to the editor for 
comments and suggestions. 
This work has been partly supported by 
Ministero dell'Universit\`a e della Ricerca Scientifica e Tecnologica 
(MURST).  

\clearpage

\begin{deluxetable}{ccccc}
\footnotesize
\tablecaption{Galaxy density fits. \label{tbl-1}}
\tablewidth{0pt}
\tablehead{
\multicolumn{5}{c}{de Vaucouleurs profile} \nl
\colhead{$\sigma_0$} & \colhead{${\rm r_v}$}   & \colhead{$\gamma_g$}   &
\colhead{$\sigma_b$} & \colhead{$P_{KS}(>D)$}\nl
\colhead{(${\rm kpc^{-2}}$)} & \colhead{(${\rm kpc}$)} & \colhead{ } & \colhead{(${\rm deg^{-2}}$)} \nl
} 
\startdata
(0.67$\pm$0.3)$\cdot 10^{-1}$ & 259$\pm$104 & 0.16$\pm$0.03 & 17$\pm$10 & 0.70 \nl
\tableline
\nl
\multicolumn{5}{c}{King profile} \nl
{$\sigma_0$} & {${\rm r_c}$}   & {$\kappa$}  &
{$\sigma_b$} & {$P_{KS}(>D)$}\nl
{(${\rm kpc^{-2}}$)} &{(${\rm kpc}$)} &{ } &{(${\rm deg^{-2}}$)} \nl
\tableline
(0.43$\pm$0.19)$\cdot 10^{-3}$ & 39$\pm$15 & 0.690$\pm$0.14 & 19$\pm$11 & 0.79 \nl
\enddata

\end{deluxetable}

\clearpage
\begin{deluxetable}{cccccc}
\footnotesize
\tablecaption{X-ray brightness fits. \label{tbl-2}}
\tablewidth{0pt}
\tablehead{
\colhead{$Image$} &
\colhead{$I_o \times 10^4$} &
\colhead{$n_o \times 10^3$} &
\colhead{$r_c$}&
\colhead{$\beta$} &
\colhead{$\chi^2/\nu$}\nl 
\colhead{ } &
\colhead{$cts \; s^{-1} arcmin^{-2}$} &
\colhead{$cm^{-3}$} &
\colhead{$Mpc$} &
\colhead{ }&
\colhead{ }\nl
} 
\startdata
Overall\tablenotemark{a} & 801$\pm$144 & 5.28$\pm$0.70 & 0.213$\pm$0.090 & 
0.518$\pm$0.070 & 174/62 \\
Southern\tablenotemark{a} & 950$\pm$171  & 6.62$\pm$0.84 & 0.159$\pm$0.071 & 
0.511$\pm$0.069 & 82/62 \\
\enddata
\tablenotetext{a}{For both images exposure time is 4715.4 s and the background is $
I_b=8.86\cdot 10^{-4}\;\; cts \; s^{-1} arcmin^{-2}$}.

\end{deluxetable}

\clearpage

\figcaption[fig1.eps]{Radial profile of the surface density of galaxies. 
Vertical bars represent the standard deviation of the number counts in the 
relevant annulus.
The curves represent the fits of unbinned data.
{\it Solid line:} King profile; {\it dotted line:} de Vaucouleurs profile.
\label{fig1}}

\figcaption[fig2.eps]{ROSAT/PSPC image of A~2319. The image is
filtered with a gaussian of 100 arcsec (FWHM). The contour levels,
starting from 0.40 counts pixel$^{-1}$ ($1 pixel = 15" \times 15"$)
increase by a factor $2^{1/2}$.
Notice the asymmetry in the N-NW
direction, due to the B component. \label{fig2}}

\figcaption[fig3.eps]{X-ray surface brightness profile of A~2319~A, after the
exclusion of the northern half of the image. Errors represent the 
standard deviation of photon counts. The solid line is the fit with
the $\beta$-model. \label{fig3}}

\figcaption[fig4.eps]{The gas density $\rho_{gas}(r)$, derived from numerical
 inversion of the projection equation (see CNT), versus the fitted 
galaxy density
$\rho_{gal}(r)$.  Vertical lines  show the relevant radii, indicated in
Mpc ($h_{50}=1$). The local slope of the curve defines the function
$\beta_{XO}(r)$.  \label{fig4}}

\figcaption[fig5.eps]{$\beta_{spec}$ versus $(\beta_{XO}+\beta_{TO})$ computed at
the radii of 4, 10, 20 arcmin, corresponding to the boundaries of the
four rings whose temperatures are given by Markevitch
(1996). \label{fig5}}

\figcaption[fig6.eps]{Temperature versus radius of A~2319~A, as deduced from
Markevitch (1996).  {\it dotted:} best fit polytropic model; {\it
short dash:} quadratic interpolation $T(r)=T_o+a r^2$. \label{fig6}}

\figcaption[fig7.eps]{The mass as a function of radius for A~2319~A. {\it
Solid:} total binding mass $M_{tot}^{isot}(r)$, for the isothermal
``$\beta$''-model with $T_{isot}=10\pm 0.7 keV$; {\it dotted line:}
$M_{tot}(r)$ for the polytropic model; {\it shaded area:} $M_{tot}(r)$
for $T(r)=T_o-ar^2$; {\it long dash:} galaxy mass $M_{gal}(r)$; {\it
dot dash:} gas mass $M_{gas}(r)$. Shaded areas represent the $1\sigma$
uncertainty as deduced by Monte Carlo simulations.  \label{fig7}}

\figcaption[fig8.eps]{Baryon fraction $f_b(r)$ as a function of radius. {\it
Solid:} isothermal ``$\beta$-model''; {\it dotted:} polytropic model;
{\it shaded area:} model with $T(r)=T_o-ar^2$ with the $1\sigma$ uncertainty as deduced
by Monte Carlo simulations.  \label{fig8}}

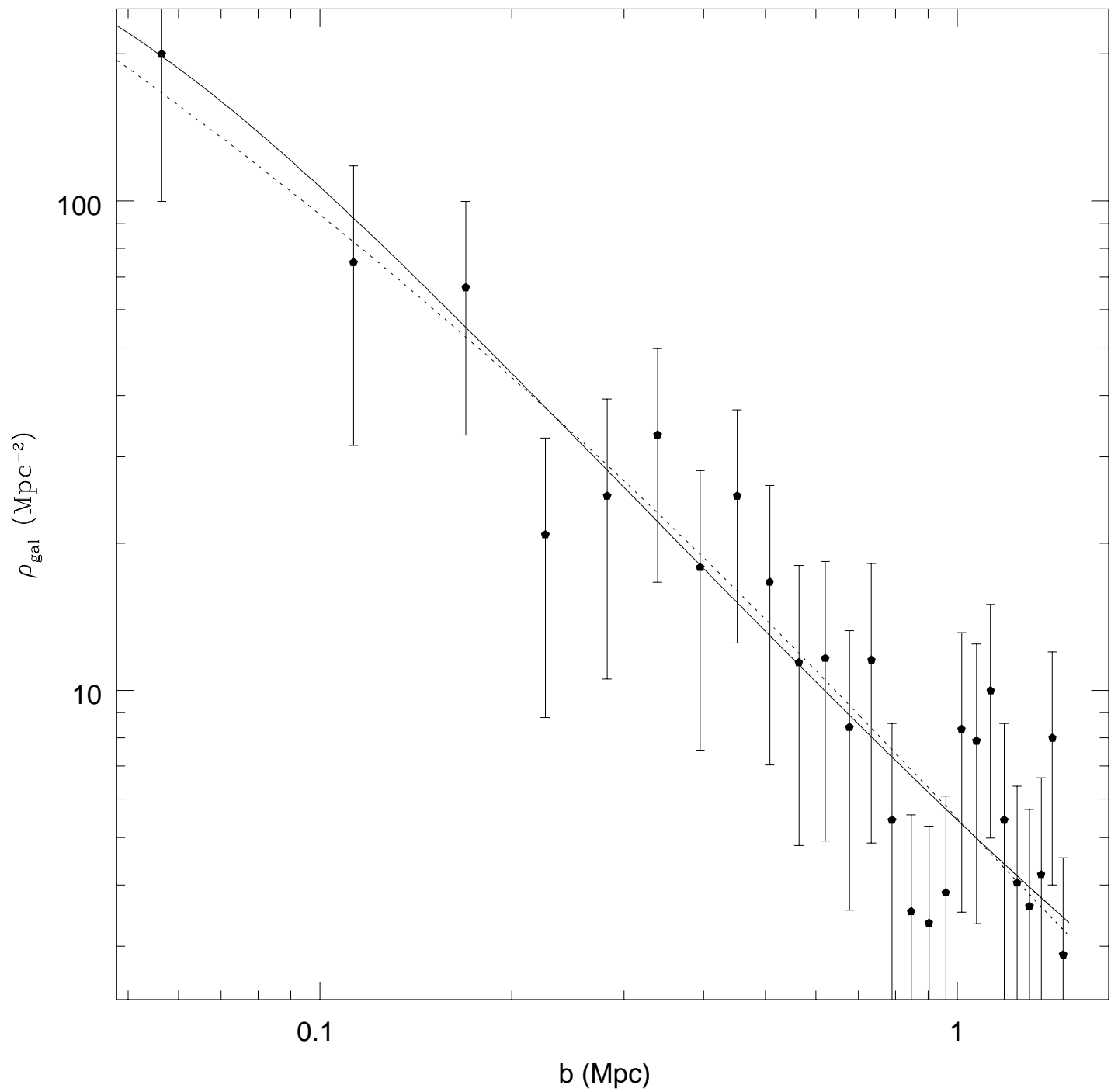
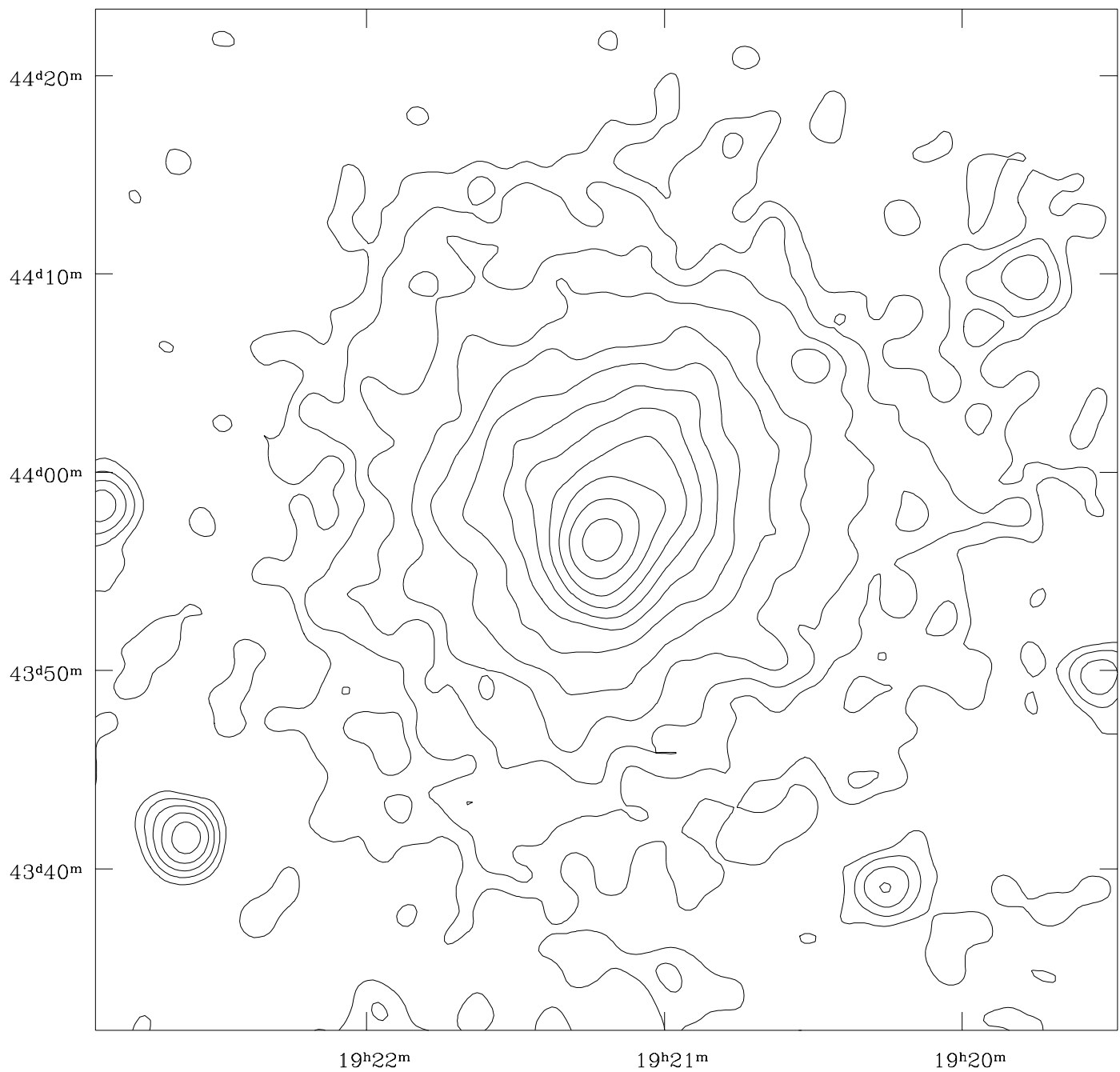
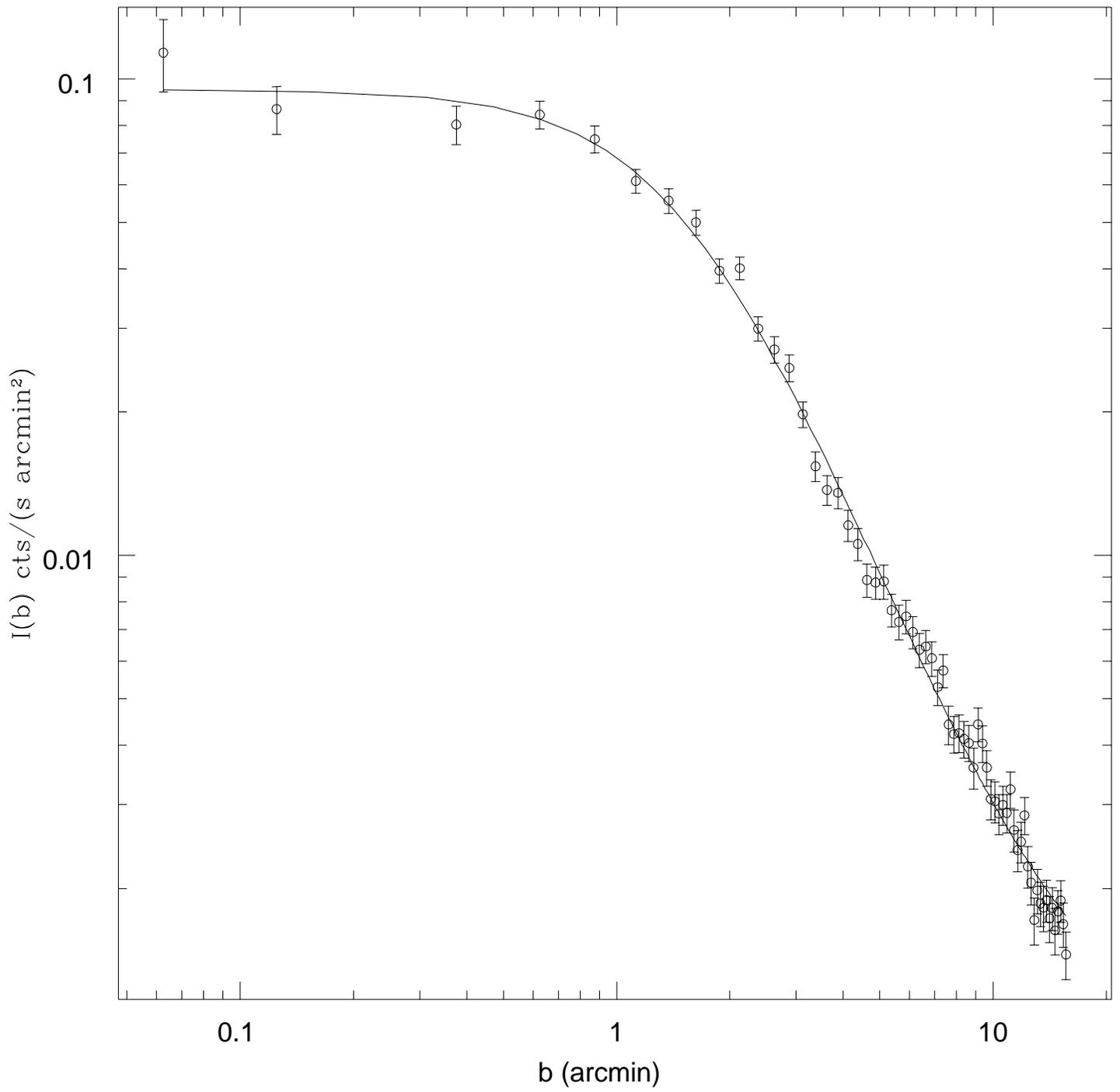
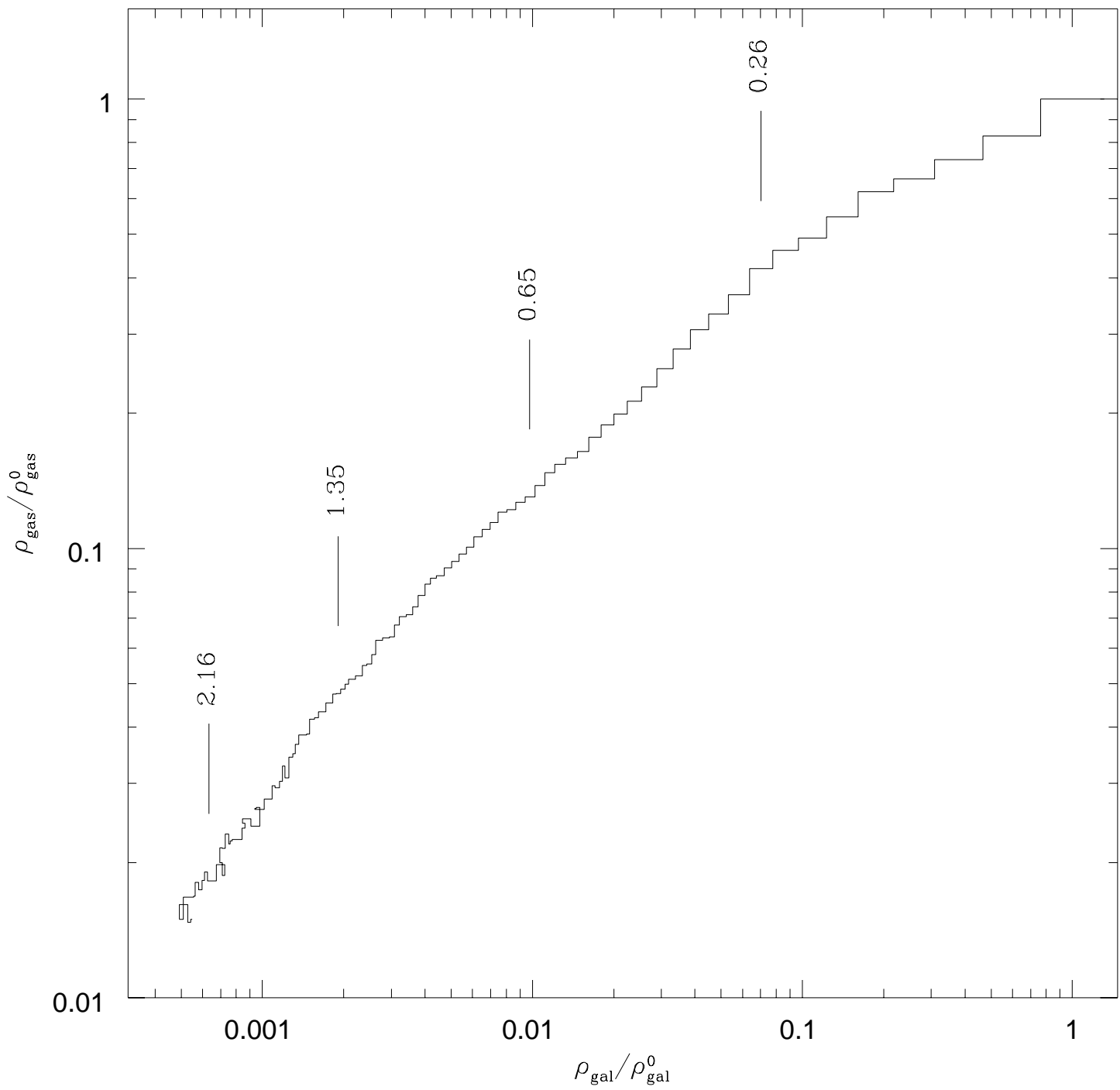
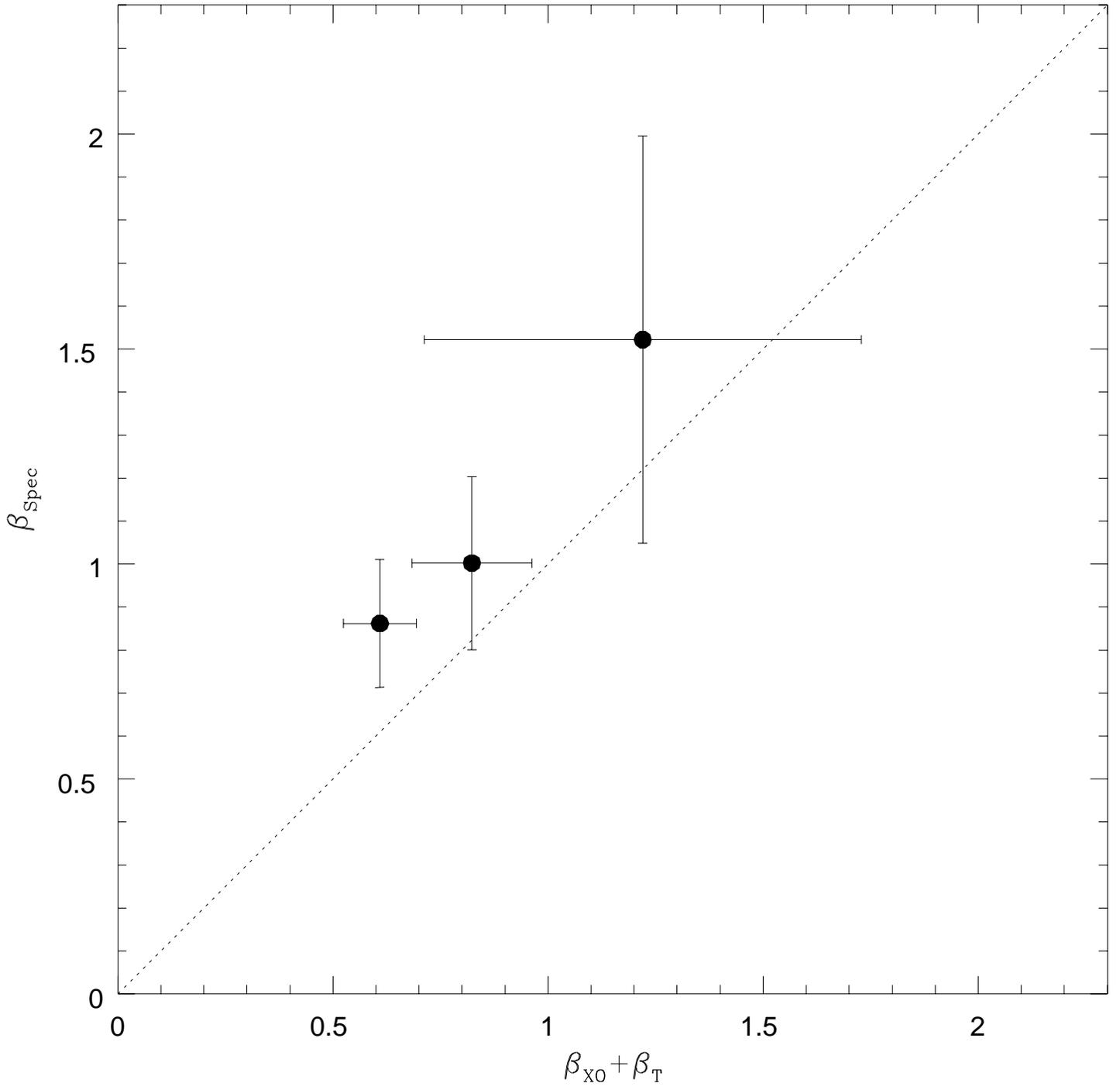
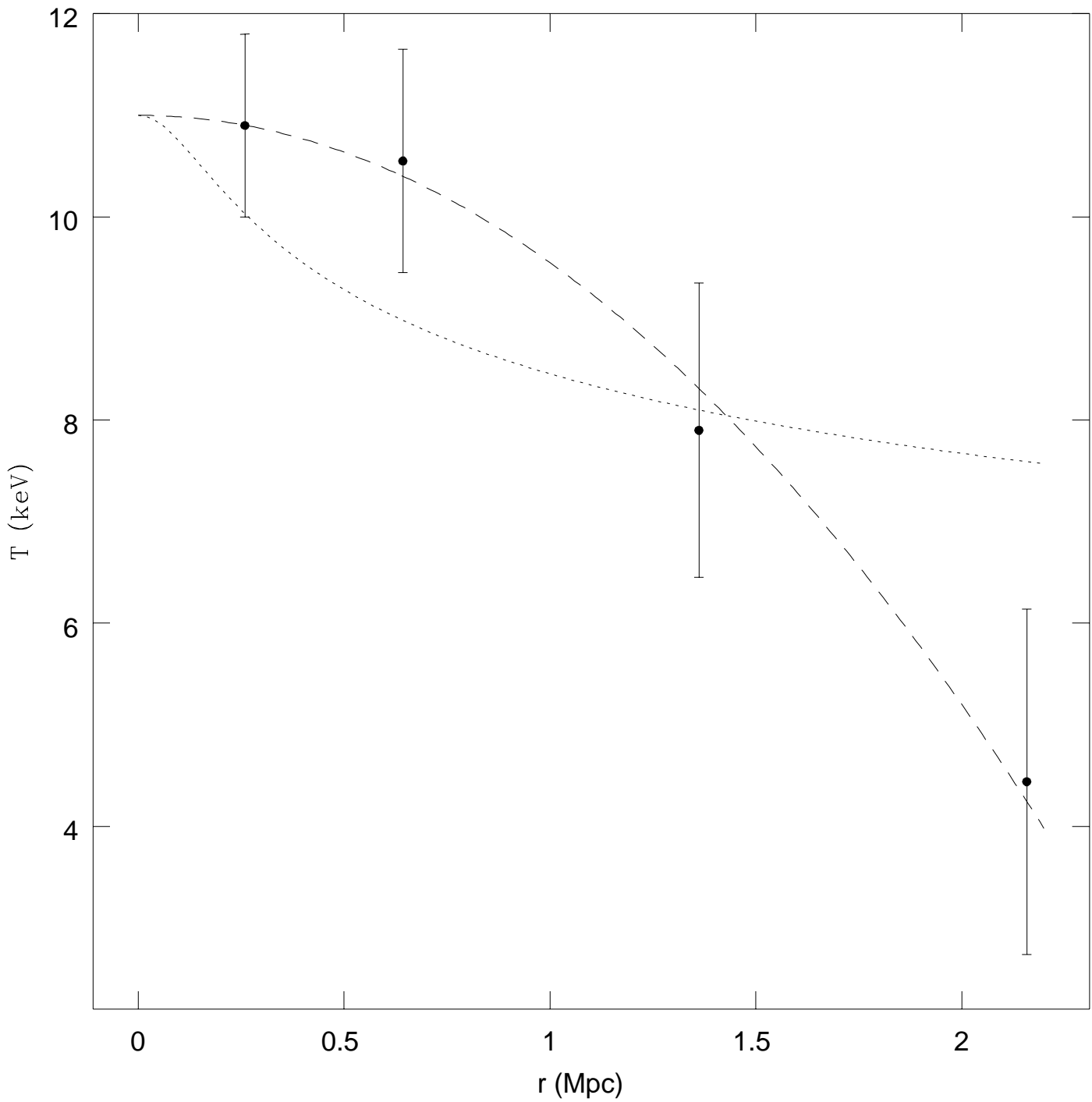
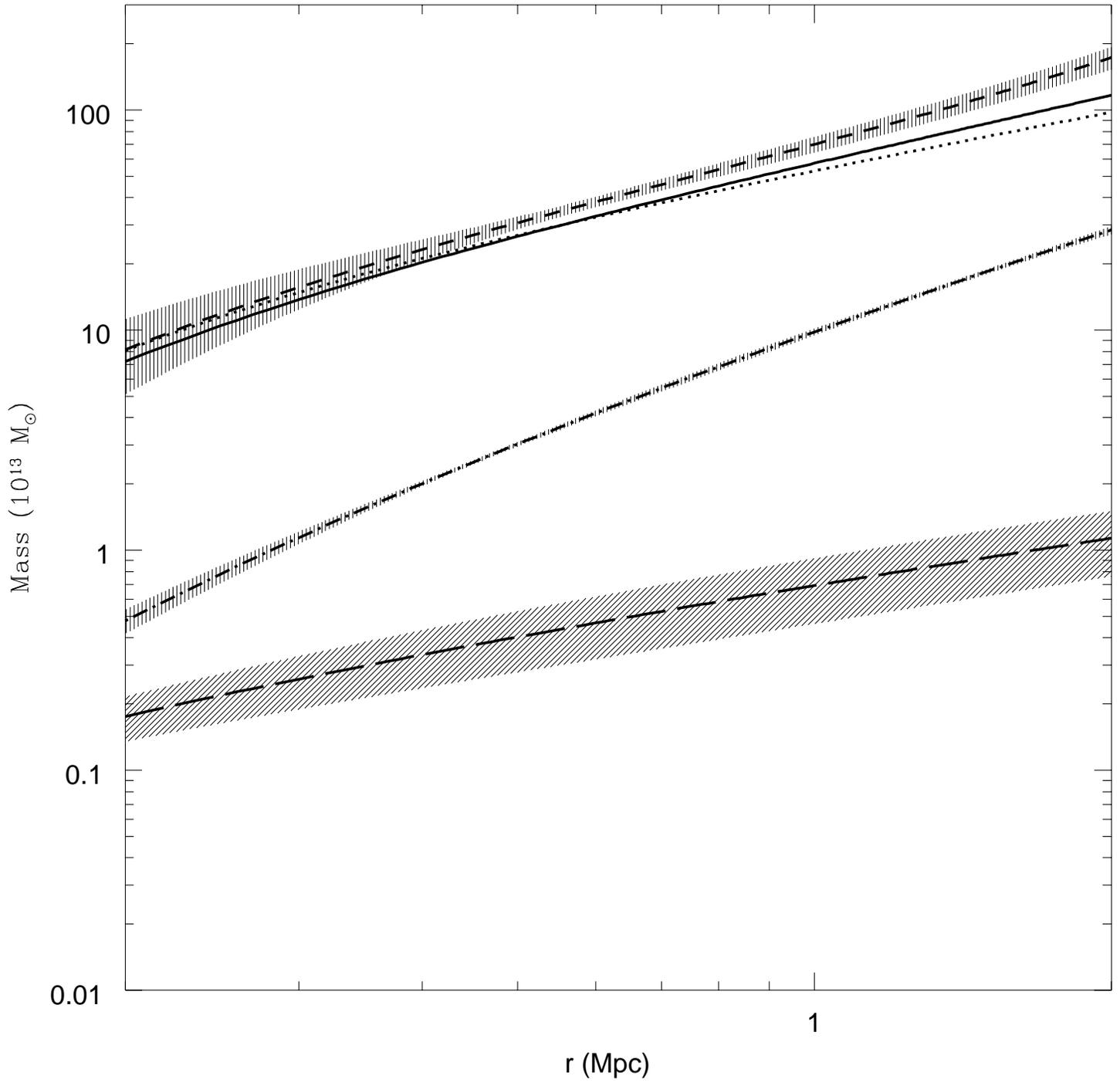
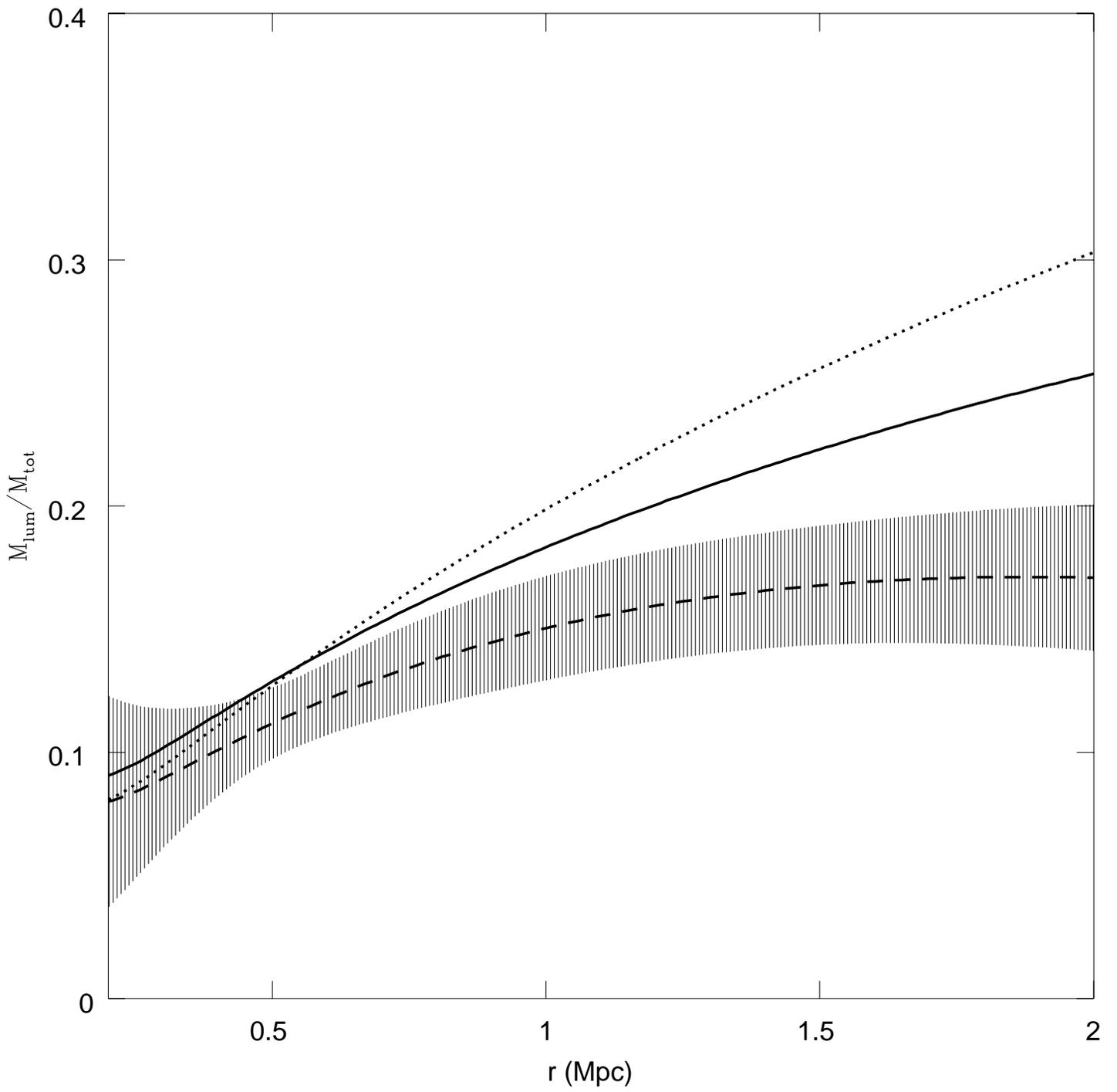
\end{document}